# Performance of the Fully Digital FPGA-based Front-End Electronics for the GALILEO Array

D. Barrientos, M. Bellato, D. Bazzacco, D. Bortolato, P. Cocconi, A. Gadea,

V. González, *Senior Member, IEEE*, M. Gulmini, R. Isocrate, D. Mengoni, A. Pullia, *Member, IEEE*,

F. Recchia, D. Rosso, E. Sanchis, *Member, IEEE*, N. Toniolo, C. A. Ur and J. J. Valiente-Dobón

*Abstract–In this work we present the architecture and results of a fully digital Front End Electronics (FEE) read out system developed for the GALILEO array. The FEE system, developed in collaboration with the Advanced Gamma Tracking Array (AGATA) collaboration, is composed of three main blocks: preamplifiers, digitizers and preprocessing electronics. The slow control system contains a custom Linux driver, a dynamic library and a server implementing network services. The digital processing of the data from the GALILEO germanium detectors has demonstrated the capability to achieve an energy resolution of 1.53‰ at an energy of 1.33 MeV.*

## I. Introduction

PART of contemporary nuclear physics research focuses on the study of the fundamental properties of exotic nuclei far from stability by means of gamma-ray detection. During the last decades, a number of large gamma-ray spectrometers have been developed in accelerator facilities, with high sensitivity and absolute efficiency capabilities. In addition, the fore coming operation of several new Radioactive-Ion Beam (RIB) facilities puts new requirements for operation with gamma-ray spectrometers while opening new scenarios for future nuclear physics experiments.

GALILEO [1] is a $4\pi$ gamma-ray spectrometer with a photopeak efficiency of 8% and a peak-to-total ratio of about 50%. The array is composed of High-Purity Germanium (HPGe) detectors surrounded by Bismuth Germanium Oxide (BGO) scintillators in order to implement Compton suppression techniques [2] for increased sensitivity. GALILEO plans to operate with stable beams provided by the Tandem-ALPI accelerator complex as well as the SPES [3] RIB facility. In order to fully exploit the capabilities of the array, it will be coupled with several ancillary detectors such as TRACE, EUCLIDES, large volume $LaBr_3$ detectors, RFD or Neutron Wall [4-5].

Detectors have been reused from previous arrays and rearranged by a custom mechanics designed in house. The structure, currently being installed at the Legnaro National Laboratories (Italy), is presented in Fig. 1. The different frames allow the installation of detectors and ease the integration of the array with other detectors in different experimental setups.

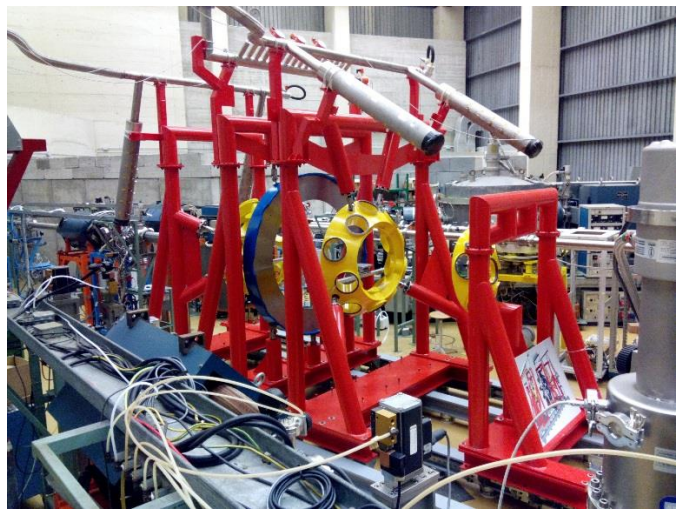

Fig. 1. Photograph of the holding structure of the GALILEO array at the Legnaro National Laboratories (Italy).

## II. Preamplifiers And Digitizers

The output signals of the HPGe detectors are simultaneously read out by advanced charge-sensitive preamplifiers employing a fast-reset technique for dead time and dynamic range optimization [6-7]. This technique consists on a fast discharge of the pole-zero capacitance when the preamplifier output signal is larger than a programmable threshold. Being the length of the saturated pulse proportional to the height of the input pulse, the dynamic range of the

Manuscript received June 16, 2014. This work has been partially supported by Istituto Nazionale di Fisica Nucleare (INFN), Italy, by the Generalitat Valenciana, Spain, under grant PROMETEO/2010/101 and by MINECO, Spain, under grants AIC-D-2011-0746, FPA2011-29854 and FPA2012-33650.

D. Barrientos, D. Bortolato, P. Cocconi, M. Gulmini, D. Rosso, N. Toniolo and J. J. Valiente-Dobón are with Istituto Nazionale di Fisica Nucleare, Laboratori Nazionali di Legnaro, viale dell'Universita' 2, 35020 Legnaro (Padova, Italy). Corresponding author: Diego Barrientos; e-mail: diego.barrientos@lnl.infn.it

M. Bellato, D. Bazzacco, R. Isocrate, D. Mengoni, F. Recchia and C. A. Ur are with Istituto Nazionale di Fisica Nucleare, Sezione di Padova, via Marzolo 8, 35131 Padova (Italy).

A. Gadea is with Instituto de Física Corpuscular (CSIC-UV), Catedrático José Beltrán 2, 46980 Paterna (Valencia, Spain).

V. González and E. Sanchis are with Departamento de Ingeniería Electrónica (Universitat de València), Escola Tècnica Superior d'Enginyeria, Avinguda de la Universitat s/n 46100 Burjassot (Valencia, Spain).

Alberto Pullia is with Dipartimento di Fisica (Universita' di Milano) and Istituto Nazionale di Fisica Nucleare Sezione di Milano, via Celoria 16, 20133 Milano (Italy).

preamplifiers thus get increased by employing the Time over Threshold (ToT) technique [8].

The differential output of the preamplifiers and the differential signals coming from the BGO scintillators are arranged into 6-way cables connected by MDR connectors at the input of the digitizer modules. The digitizer of GALILEO uses the boards developed by our group for the second generation of electronics for the Advanced Gamma Tracking Array (AGATA) [9], and is composed of six Digi-Opt12 boards [10] and two Control Cards [11-12] assembled as shown in Fig. 2.

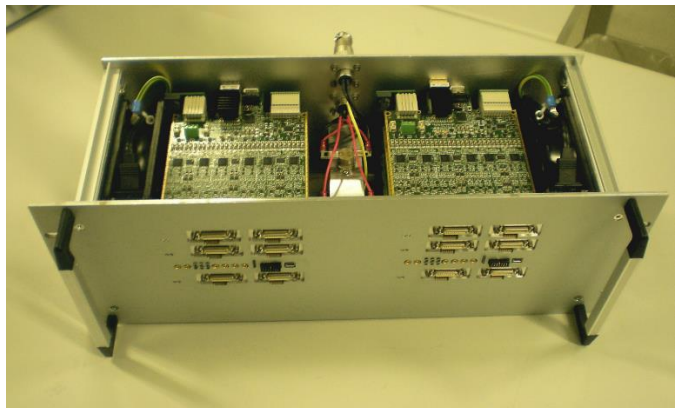

Fig. 2. Photograph of the digitizers module. Three Digi-Opt12 boards and one Control Card are organized in two columns. The boards are cooled from two fans located at the outer sides of the box.

The Digi-Opt12 board is a low power 12-channel digitizer board with optical output that performs a digitization of the differential input signals at 100 Msps with 14 bits of resolution. The digital output is sent to a 12-fiber optical cable by means of high-speed serial links, encoded with the JESD204A protocol [13] and working at 2 Gbps each. Each channel has two programmable input ranges, corresponding to energy ranges of 0-7 or 0-20 MeV for gamma-rays interacting in the HPGe detectors equipped with the GALILEO-type preamplifiers. In addition, the analog offset of each channel is also programmable from the slow control system and a synchronization signal can be injected at the input of the Analog-to-Digital Converters (ADC) to compensate different channel latencies due to unbalanced digital paths. The sampling clock, synchronization signal and slow control buses are provided from a custom backplane connector on the rear side of the digitizer.

The Control Card has three main tasks: receives (from the preprocessing electronics) and broadcasts (to the three associated Digi-Opt12 boards) the sampling clock and synchronization signals and performs the slow control of the associated Digi-Opt12 boards and the Control Card itself. The board is equipped with a Xilinx Spartan-6 Field Programmable Gate Array (FPGA) to fulfill these tasks. Special care has been put in the Control Card to provide sampling clock and synchronization signals with minimum contributions of jitter and skew, a task which required specific circuitry, simulations and tests. Concerning the slow control, a set of synchronized registers placed in the Control Card and preprocessing FPGAs have been linked to the physical slow control buses in order to control each device in the electronic chain from the slow control system, as described in Section IV. The implementation details of this technique are specified in [14].

The communications between the digitizer module and the preprocessing electronics are performed by optical links, whereas the boards within the digitizer module are linked by means of custom backplanes. The digitizer module is divided in two parts as each Control Card manages three Digi-Opt12 boards. The digitizer module is cooled by two fans that provide low mechanical noise and the power consumption of the whole digitizer box is less than 90 W for 72 digitizing channels, i.e. 1.25 W/channel.

### III. HARDWARE PROCESSING

The preprocessing board is a custom PCI express board that receives digital data from the digitizer module, process and routes them to the hosting PC by means of a 4x PCI express link. The data from 36 high-speed links at 2 Gbps are processed in a Xilinx Virtex-6 FPGA, where custom algorithms select the information from events of interest and send the data to the hosting PC via the PCI express link at a sustained rate up to 400 MB/s. A photograph of the preprocessing board, developed in collaboration with the AGATA project, is presented in Fig. 3.

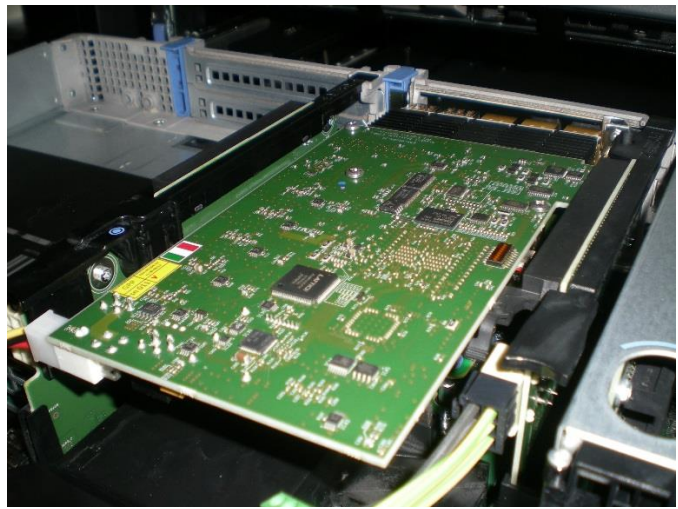

Fig. 3. Photograph of the preprocessing board mounted at the hosting PC.

In the following subsections, the main features implemented in the preprocessing board are described in detail.

#### A. Energy computation and short traces capture

After JESD204A protocol decoding, the input data are fed to an embedded First-In First-Out (FIFO) in order to compensate the different channel latencies, as discussed in the previous section. Samples then get processed in a first-level trigger generation module and in an energy computation module.

First-level trigger is fired when the corresponding energy of the input pulse exceeds a programmable threshold. A trigger activates the energy computation mechanism and the capture of the trace, e.g. recording the trace in a time window around the triggering sample. The trigger module sends a trigger request to the Global Trigger and Synchronization (GTS) system, which is in charge of the trigger coordination of the whole array and of other detectors in the setup [15-16].

When the input rate is very low, the system generates fake first-level triggers (not sent to the GTS system) that produce tagged idle events arriving at the Data Acquisition (DAQ) system [17]. The production of a user-selectable minimum event rate by means of the idle mechanism (similar to the one implemented in the GTS system [15]) simplifies dramatically the operation of the event builder and may become particularly useful when working at very low rates, as it is foreseen with the first beams of the SPES accelerator facility.

The energy of the triggered gamma-ray is computed from the input pulse using a trapezoidal shaping [18-20], tuned with a set of programmable parameters controlled by the slow control system. The energy computation module also includes an auto-triggered baseline restorer module that improves the performance of the system when working at high counting rates.

Upon a trigger request, the GTS system provides a timestamp value and eventually replies to the request within a 20 μs time window. The computed energy and trace are stored in Random Access Memory (RAM) while waiting for the validation. In case of an accepted event, data are packaged and sent to the PC, while a rejection or time out frees the RAM area for another incoming event.

### B. GTS leaf services

The preprocessing board hosts the GTS leaf services that integrate the card with the global triggered timing system.

The GTS system is in charge of the precise synchronization and trigger management of GALILEO and associated ancillary detectors. It provides a phase-aligned 100 MHz clock to its leaves (preprocessing boards) with sub-nanosecond precision. This clock is also sent through dedicated optical links to the digitizer module with deterministic latency and hence to each ADC in the system. As mentioned in the previous section, the deterministic latency is obtained by injecting a synchronization signal at the input of the ADCs, also provided synchronously by the GTS system and arriving at the digitizer module from another dedicated optical link.

The GTS leaf service implementation has been tested using a small GTS tree with a single root node implemented in a Common Mezzanine Card (CMC) and two GTS leaves implemented in preprocessing boards, as shown in Fig. 4.

### C. Long traces capture and online spectra

Trace capture upon a trigger event in preprocessing FPGAs is composed of a few hundred of samples. While enough for Pulse Shape Analysis (PSA) algorithms [21-22], the normal trace length is not adequate when, for example, analyzing the input signal in the frequency domain or trying to visualize small issues in malfunctioning channels.

In order to cope with this issue, a dedicated dual-port RAM in the preprocessing FPGA can either store the input samples to the preprocessing algorithms or perform online spectra with the computed energies.

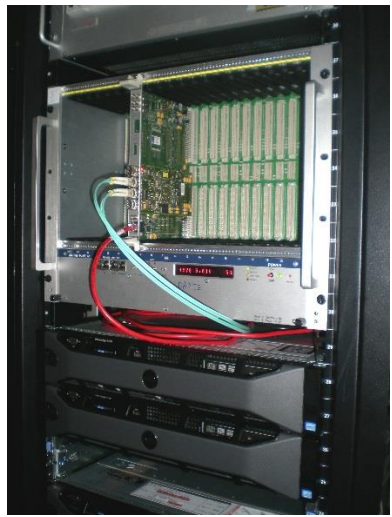

Fig. 4. Photograph of the test bench with a GTS root node, at the top side of the picture, and two leave nodes in the preprocessing boards placed at the hosting PCs. The GTS root mezzanine is placed in a VME carrier. The optical links of the GTS tree are the green fibers in the picture, whereas the red Ethernet cable is used for the slow control of the GTS mezzanine.

Longer traces may be captured in triggered or non-triggered mode, and may have a length up to a few hundreds of thousands of samples. Alternative to the readout of longer traces, the dual-port RAM can be used to histogram in real-time the detector pulse amplitudes. These histograms are incremented irrespective of the GTS trigger validation, thereby providing the true single spectra seen by the detectors.

## IV. SLOW CONTROL SYSTEM

In order to manage the control of the whole electronic chain, a set of registers in the preprocessing FPGA is mapped to the user-memory of the hosting server by means of a custom driver. The developed driver has been successfully integrated in different Linux distributions and kernel versions.

The driver creates virtual devices (in the user-space of the operating system) that map the internal registers of the FPGA and can be easily accessed using read and write operations. The content of the registers varies from parameters of the shaping filter for energy computation to values of physical slow control buses in real-time.

A C++ dynamic library encapsulates the low level operations of the different hardware-dependent devices and provides a user-friendly Application Programming Interface (API) to higher-level users. The library encapsulates also the communications to digitizer modules, as the devices are remotely controlled by a set of synchronous registers described in Section II. The API allows identical operations in local or remote devices in a fully transparent way.

The dynamic library is instantiated by a server that operates several state-machines taking care of the state of the hardware underneath. The server implements Web Services Description Language (WSDL) network services [23] to provide an interface for the control the state-machines.

The WSDL clients connecting to different servers in the system may be single or multiple Graphical User Interfaces (GUI) or more complex control system, for example, distributed diagnostics.

## V. SYSTEM PERFORMANCE

The first set of cards presented in this work has been already tested with GALILEO germanium detectors at the Legnaro National Laboratories (LNL). A picture of the detectors used in the test bench is presented in Fig. 5.

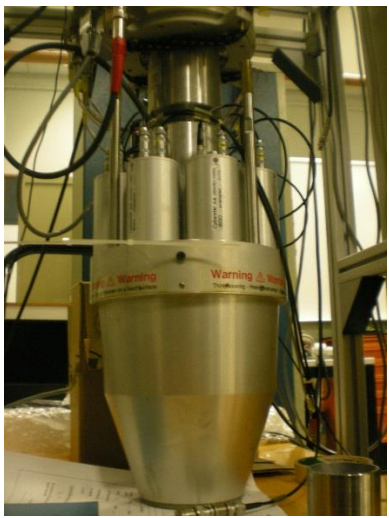

Fig. 5. HPGe GALILEO detector with anti-Compton shield used for the tests of the digital electronics chain.

The digital read out using the new setup has shown equal or better performance when compared to conventional analog electronics. An energy spectrum obtained with a GALILEO HPGe detector with its digital acquisition system is presented in Fig. 6. The detector has been illuminated with $^{241}$Am and $^{60}$Co gamma-ray sources in order to quantify the energy resolution at low and high energies in the range of interest. The Full Width at Half Maximum (FWHM) of the peaks at 59.6 keV and 1332.5 keV in the spectrum is 0.972 keV and 2.035 keV respectively. Therefore, the resolution at low energy is about 1.63% and about 1.53‰ at high energy.

## VI. CONCLUSIONS AND FUTURE WORK

The cards presented have been prototyped, manufactured and qualified in a test bench that is in operation at the experiment site. The fully-digital read out of the GALILEO array has been described in terms of its constituting items. Some features such as energy computation, GTS integration, traces and spectra capture, as well as the complex slow control system and system performance have been described in detail.

Integration of the boards with the present electronics of the AGATA array and the first tests of the GALILEO array operated with stable beams are scheduled to take place in the next few months.

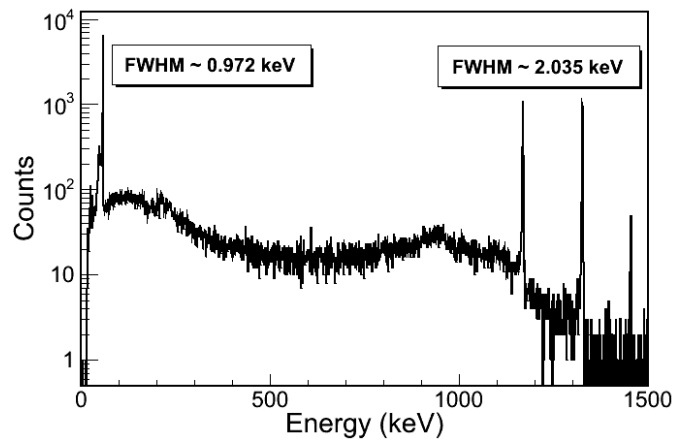

Fig. 6. Energy spectrum obtained with a GALILEO germanium detector illuminated with $^{241}$Am and $^{60}$Co gamma-ray sources.